\documentclass[a4paper,11pt]{article}
\usepackage{subcaption}
\usepackage{jheppub} 
\usepackage{lineno}
\usepackage{physics}
\usepackage{booktabs}
\usepackage{multirow}

\usepackage{xspace}

\usepackage[T1]{fontenc}
\usepackage[scaled=0.85]{beramono}

\usepackage{mathrsfs}

\title{\boldmath Production Shape Effects Beyond Rate Rescaling in Dark Photon Searches}

\author[a]{X. Zhong}
\author[]{,}
\author[a]{H.Y. Qiu}
\author[]{,}
\author[a]{X.Q. Shen}
\affiliation[a]{School of Physics and Mechanical and Electrical Engineering, Longyan University, \\
Longyan 364012, China}

\emailAdd{15521018891@163.com}

\abstract{
Production model uncertainties in long-lived particle searches can affect
both the inclusive yield and the kinematic distributions sampled by the
experimental acceptance. We isolate the latter effect by comparing the modified
Weizs\"acker--Williams description with the effective quasireal approximation
for dark photon production via proton bremsstrahlung after normalizing both
spectra on a common physical domain.
 At fixed production yield within this domain, the resulting acceptance difference directly measures the departure from rate
factorization. In the NA62 beam dump geometry, the central calculation gives a fractional acceptance shift
$\delta_{\rm NF}>0.3$ at every numerically converged grid point
within the displayed NA62 sensitivity overlap. This conclusion persists for effective QRA off-shell scales of
1.0, 1.5, and 2.0~GeV and under the aperture variations considered.
Separately, the public NA62 full selection acceptances give a
nonconstant dimuon acceptance ratio between the Mixing and
Bremsstrahlung production categories,
ranging from approximately
0.58 to 4.23, with a median absolute departure from unity of
approximately 0.79 over the analyzed HEPData grid. This provides
complementary evidence that the published full selection acceptance
depends on the production category. More generally, the projection criterion provides a quantitative test of rate
factorization. The normalized production shape variation must have negligible
projection onto the acceptance response at the required accuracy. When this
condition is not satisfied, production model dependence must be propagated
through the acceptance calculation or treated as an acceptance uncertainty.
}

\begin{document}
\maketitle
\flushbottom

\section{Introduction}

Searches for long-lived and feebly interacting particles are
intrinsically differential. The expected signal yield depends not
only on the inclusive production rate but also on the energy and
angular distributions that determine whether a particle reaches the
fiducial volume, decays within it, and produces reconstructible
decay products. This kinematic dependence is central to the design
and interpretation of beam dump and far forward experiments
\cite{r1,r2,r3,r4,r5,r6,r7,r8,r9,r10}. Modern recasting and sensitivity frameworks therefore propagate production kinematics through geometry, decay probability, reconstruction, and selection \cite{r11,r12,r13,r14,r15,r16,r17}. This raises the question of whether differences between alternative
descriptions of a production mechanism can be reduced to a rate
rescaling after their production rates are matched on a common
physical domain.

Dark photon production by proton bremsstrahlung provides a controlled setting
for this question. Predictions have progressed from early fixed target and beam
dump treatments to forward hadronic descriptions with improved splitting
functions and timelike proton form factors~\cite{r18,r19,r20,r21,r22}. Recent
work has further emphasized that hadronic uncertainties affect both total rates
and differential spectra~\cite{r22,r23,r24,r25,r26}. In conventional sensitivity
comparisons these effects enter simultaneously. Consequently, a difference in
the predicted sensitivity does not by itself determine whether the underlying
production model dependence can be represented by a normalization uncertainty.
Separating the production rate from the normalized production shape is therefore
necessary for a direct test of rate factorization.

We address this issue by comparing the modified Weizs\"acker-Williams (mWW)
description with the effective form of the quasireal approximation (QRA) on a common
physical domain. The two production spectra are normalized before the same
acceptance response is applied. This construction removes the production rate
difference within the common domain and isolates the acceptance change induced
by the normalized production shape. The resulting acceptance difference
therefore provides a direct test of rate factorization at fixed production yield.
More generally, the acceptance shift is determined by the projection of the
normalized production shape variation onto the acceptance kernel. This
projection provides a quantitative criterion for determining when production
model dependence can be represented by a single rate normalization.

We test this criterion for dark photons in the NA62 beam dump
geometry~\cite{r27,r28,r29,r30}. In the overlap between the published NA62 sensitivity region for the BC1 dark photon benchmark and the parameter space covered by the fiducial calculation, all
central grid points retained by the numerical convergence criterion
show an acceptance shift above 30\%, and this conclusion survives
variations of the effective QRA off-shell scale $\Lambda_p$ and of
the aperture considered below. 

Public NA62 acceptances at fixed visible final state provide a complementary
test at the selection level through a nonconstant dimuon acceptance ratio
between the Mixing and Bremsstrahlung production categories. This observable
does not directly test the mWW to effective QRA comparison and is not used
to construct a modified exclusion. Instead, it independently shows that the
published full selection acceptance depends on the production category. Together, these results quantify a general limitation of rate-only long-lived particle recasting, while the projection criterion provides
a practical diagnostic for deciding when production model dependence
must be retained in the acceptance calculation.

\section{Production and acceptance nonfactorization}
\label{sec:nonfactorization}

Hadronic production affects a search for particles with macroscopic lifetimes through both the total number of particles produced and the kinematic distribution on which the acceptance acts. A rate-only replacement changes the production normalization while retaining the acceptance of a baseline model. This procedure is valid only when the normalized production shape does not appreciably change the accepted fraction.

For the azimuthally symmetric production models considered here, with $M\in\{\mathrm{mWW},\mathrm{QRA}\}$, let \(\Phi\) denote the nonazimuthal production variables that determine the dark photon momentum and polar kinematics, while \(\phi\) denotes the azimuthal angle about the beam axis. At fixed dark photon mass \(m_{A^\prime}\), the azimuth-integrated differential yield on the common physical domain \(\mathcal{D}(m_{A^\prime})\) is written as
\begin{equation}
\label{eq:production_shape}
\begin{gathered}
\frac{dN_M}{d\Phi}
=
N_M^{\mathcal{D}}(m_{A^\prime})
f_M(\Phi,m_{A^\prime}),
\\
\int_{\mathcal{D}(m_{A^\prime})}
d\Phi\,
f_M(\Phi,m_{A^\prime})
=
1 .
\end{gathered}
\end{equation}
The normalization $N_M^{\mathcal{D}}$ fixes the production rate in the
common domain, while $f_M$ fixes the distribution in the
variables that enter the acceptance. Normalizing both
models on the same domain removes the production rate difference
within that domain from the acceptance comparison.

The contribution of geometry and decay to the accepted fraction is
written as the average of a common acceptance kernel over the
normalized production spectrum,
\begin{equation}
\label{eq:geom_acceptance}
A_{\mathrm{geom}}^{M}(m_{A^\prime},c\tau)
=
\int_{\mathcal{D}(m_{A^\prime})}
d\Phi\,
f_M(\Phi,m_{A^\prime})
K(\Phi,m_{A^\prime},c\tau) .
\end{equation}

The aperture can vary with the decay position, so the kernel is evaluated along the particle trajectory,
\begin{equation}
\begin{aligned}
K(\Phi,m_{A^\prime},c\tau)
&=
\int_{\mathrm{FV}} ds\,
G(\Phi,s)
\frac{\exp[-s/\lambda(\Phi,m_{A^\prime}, c\tau)]}
{\lambda(\Phi,m_{A^\prime}, c\tau)},
\\
\lambda(\Phi,m_{A^\prime}, c\tau)
&=
\beta\gamma(\Phi,m_{A^\prime})c\tau .
\end{aligned}
\label{eq:acceptance_kernel}
\end{equation}
Here $s$ is the path length measured from the TAX reference position,
$c\tau$ is the proper decay length, and $\beta\gamma$ is the Lorentz
boost factor. Production is approximated by a fixed point source at
the NA62 TAX beam dump reference position, common to both production
models, with ${\rm FV}$ denoting the fiducial decay region. The position dependent geometric weight is
\begin{equation}
\label{eq:geometry_weight}
G(\Phi,s)
=
\frac{1}{2\pi}
\int_{0}^{2\pi}
d\phi\,
\chi_{\mathrm{ap}}(\Phi,\phi,s) ,
\end{equation}
where \(\chi_{\mathrm{ap}}(\Phi,\phi,s)\) equals one when the trajectory at that decay position satisfies the local aperture condition $r_\perp(\Phi,\phi,s)<R(Z(s))$ and equals zero otherwise. 
Here $r_\perp(\Phi,\phi,s)$ denotes the transverse distance from the
beam axis, $R(Z)$ the local aperture radius, and $Z(s)$ the
longitudinal coordinate of the trajectory at path position $s$.
The geometric weight and the decay density are therefore evaluated at the same point along the trajectory. The kernel is common to both production models, while the model dependence enters through the normalized production distribution.

The shape difference between effective QRA and mWW production
can be written as
\begin{equation}
\label{eq:shape_difference}
\begin{gathered}
\Delta f(\Phi,m_{A^\prime})
=
f_{\mathrm{QRA}}(\Phi,m_{A^\prime})
-
f_{\mathrm{mWW}}(\Phi,m_{A^\prime}),
\\
\int_{\mathcal{D}(m_{A^\prime})}
d\Phi\,
\Delta f(\Phi,m_{A^\prime})
=
0 .
\end{gathered}
\end{equation}
At fixed \(m_{A^\prime}\) and \(c\tau\), the resulting acceptance shift is therefore
\begin{equation}
A_{\mathrm{geom}}^{\mathrm{QRA}}
-
A_{\mathrm{geom}}^{\mathrm{mWW}}
=
\int_{\mathcal{D}(m_{A^\prime})}
d\Phi\,
\Delta f(\Phi,m_{A^\prime})
K(\Phi,m_{A^\prime},c\tau) .
\label{eq:acceptance_shift}
\end{equation}
This form isolates the change caused by the normalized production shape and does not require a pointwise ratio of the two spectra.

At this level, a rate-only replacement predicts
\(N_{\mathrm{acc}}^{\mathrm{QRA,rate}}
=
N_{\mathrm{QRA}}^{\mathcal{D}}
A_{\mathrm{geom}}^{\mathrm{mWW}}\),
whereas retaining the QRA production shape gives
\(N_{\mathrm{acc}}^{\mathrm{QRA}}
=
N_{\mathrm{QRA}}^{\mathcal{D}}
A_{\mathrm{geom}}^{\mathrm{QRA}}\).
Their ratio is therefore
\begin{equation}
\begin{aligned}
F_{\mathrm{NF}}(m_{A^\prime},c\tau)
&\equiv
\frac{N_{\mathrm{acc}}^{\mathrm{QRA}}}
{N_{\mathrm{acc}}^{\mathrm{QRA,rate}}}
=
\frac{A_{\mathrm{geom}}^{\mathrm{QRA}}}
{A_{\mathrm{geom}}^{\mathrm{mWW}}},
\\
\delta_{\mathrm{NF}}
&\equiv
F_{\mathrm{NF}}-1 .
\end{aligned}
\label{eq:fnf}
\end{equation}
Values of \(F_{\mathrm{NF}}\) close to one mean that the normalized shape change produces little net change in the geometry and decay acceptance for this model comparison. They do not establish independence of the acceptance from arbitrary production shape variations. A significant deviation from one shows that the production shape changes the accepted fraction and invalidates a rate rescaling that keeps the baseline mWW geometry and decay acceptance fixed.
The quantity \(A_{\mathrm{geom}}\) is not a detector efficiency. It excludes trigger response, reconstruction, particle identification, and final analysis selections, and isolates the part of the signal probability fixed by production kinematics, geometry, and decay position. Both production models are normalized on the same physical domain
and are passed through the same acceptance kernel using the same
fiducial geometry. A deviation of \(F_{\mathrm{NF}}\) from one therefore measures an acceptance change generated by the difference between the normalized production shapes rather than a hidden normalization difference.

Equations~\eqref{eq:acceptance_shift} and~\eqref{eq:fnf} also provide a general criterion for rate factorization. For a normalized production shape
variation $\Delta f$, the fractional acceptance shift relative to
the rate-only baseline is $\delta_{\rm NF}$. Requiring its magnitude
to be no larger than a tolerance $\eta$ gives
\begin{equation}
\left|
\frac{
\int_{\mathcal{D}(m_{A^\prime})} d\Phi\,
\Delta f(\Phi,m_{A'})K(\Phi,m_{A'},c\tau)
}{
A_{\rm geom}^{\rm mWW}(m_{A'},c\tau)
}
\right|
=
|\delta_{\rm NF}|
\lesssim \eta .
\label{eq:factorization_criterion}
\end{equation}
Thus, production model dependence can be compressed into a single
rate normalization within the common domain only when the normalized
shape variation has negligible projection onto the acceptance kernel
at the required accuracy. Otherwise, the production dependence should be retained in the acceptance calculation or be treated as an acceptance uncertainty. The numerical implementation of the mWW and effective QRA comparison
on the common domain is summarized in Appendix~\ref{sec:numerical_implementation}.

For a complementary test at the selection level, we use the public NA62
HEPData full selection acceptances for the dimuon (2Mu) final state.
These acceptances are conditional on the dark photon reaching the fiducial
volume and decaying therein and therefore probe reconstruction and analysis
selections after this fiducial condition has been imposed. We define
\begin{equation}
F^{\rm sel}_{\rm NF,\mu\mu}
=
\frac{
A^{\rm sel}_{\rm Mixing,2Mu}
}{
A^{\rm sel}_{\rm Brems,2Mu}
}.
\label{eq:fullselection-ratio}
\end{equation}
Further details of the point-by-point construction from the published
HEPData tables are given in Appendix~\ref{sec:OF_appendix}.

\section{Results}
\label{sec:results}

\begin{figure}[t]
\centering
\includegraphics[width=0.86\columnwidth]{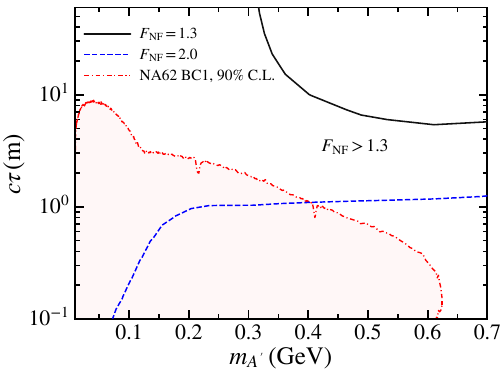}
\caption{
Central fiducial production--acceptance nonfactorization in the $(m_{A'},c\tau)$ plane for $\Lambda_p=1.5~\mathrm{GeV}$.
The black solid and blue dashed curves denote $F_{\rm NF}=1.3$ and $F_{\rm NF}=2.0$, respectively. The red dash-dotted boundary and light-red shaded region show the NA62 BC1 90\% C.L. domain after conversion to $(m_{A'},c\tau)$.}
\label{fig:fnf}
\end{figure}

The central calculation shows that the acceptance ratio departs from
unity over an extended set of retained points in the
$(m_{A'},c\tau)$ plane. To compare this behavior with published NA62
sensitivity, the BC1 90\% C.L. exclusion contour is converted point
by point from $(m_{A'},\epsilon)$ to $(m_{A'},c\tau)$ using
$c\tau=\hbar c/\Gamma_{A'}(m_{A'},\epsilon)$. Here $\epsilon$ denotes the dark photon mixing parameter, and
$\Gamma_{A'}$ is evaluated with the ALPINIST dark photon decay width implementation~\cite{r15}, including the kinematically open leptonic channels and the
hadronic contribution obtained from its tabulated hadronic
$R$-ratio input. The resulting overlap in the $(m_{A'},c\tau)$ plane is shown in
Fig.~\ref{fig:fnf}. All retained grid points in the displayed overlap
satisfy $\delta_{\rm NF}>0.3$ for the central
$\Lambda_p=1.5~\mathrm{GeV}$ calculation.

The fiducial calculation shows that the model dependence cannot be represented
by a single rate normalization on the common domain. Even after the production
rate difference has been removed by construction, the effective QRA production
shape changes the geometry and decay acceptance by more than 30\% relative to
the mWW baseline at every retained point in the displayed overlap. The remaining
difference arises from the laboratory distribution of the dark photon. Events
are redistributed among angles and boosts that are sampled differently by the
NA62 aperture and by the fiducial decay probability. The accepted fraction
therefore changes even at fixed production yield within the common domain.

The displayed BC1 contour, corresponding to the published
bremsstrahlung result without the timelike form factor, is used
only to identify a region in which the NA62 analysis has sensitivity.
We use this contour to avoid introducing the additional meson mixing
contribution present in the resonant enhancement benchmark, which is
not modeled in the present fiducial calculation. Because the present
calculation does not include the full detector response, the overlap
is not interpreted as a direct recast of the BC1 exclusion or as a
quantitative estimate of its shift.

The result $\delta_{\rm NF}>0.3$ at all retained grid points in the
BC1 overlap remains unchanged for
$\Lambda_p=1.0$, $1.5$, and $2.0~\mathrm{GeV}$ and after tightening
the numerical threshold used to implement the effective QRA hierarchy
conditions from 0.20 to 0.10. Convergence of the production
phase-space integration is assessed by comparison with a
lower-resolution integration. Points for which
$\lvert F_{\rm NF}^{\rm high}-F_{\rm NF}^{\rm low}\rvert>0.03$
are excluded, where ``high'' and ``low'' denote the central and
lower-resolution integrations, respectively.
Rescaling the reconstructed aperture
profile $R(Z)$ by $\pm10\%$ at the central value
$\Lambda_p=1.5~\mathrm{GeV}$ likewise leaves all retained grid
points in the overlap above $\delta_{\rm NF}=0.3$. We therefore
take $\Lambda_p=1.5~\mathrm{GeV}$ as the central value, with the
two endpoint values of $\Lambda_p$ and the aperture rescalings treated as robustness variations. Thus, the observed
nonfactorization occurs in parameter space overlapping the displayed
NA62 BC1 sensitivity region and is not driven by a single choice of
$\Lambda_p$, aperture profile, or numerical integration resolution.

For the complementary observable defined in Eq.~\eqref{eq:fullselection-ratio}, the dimuon acceptance
ratio between the Mixing and Bremsstrahlung production categories ranges
from approximately 0.58 to 4.23 over the analyzed HEPData window, with a
median absolute departure from unity of approximately 0.79 over the analyzed
HEPData grid. Its mass and lifetime dependence is shown in Fig.~\ref{fig:fullselection}.
The departure extends across neighboring grid points over broad
portions of the sampled mass and lifetime plane rather than being
confined to isolated points.

The fiducial calculation and the public full-selection comparison
probe distinct aspects of production dependence. The former isolates
the acceptance change between the mWW and effective QRA
bremsstrahlung descriptions, whereas the latter tests how the
published full-selection acceptance depends on the production
category at fixed visible final state. Accordingly, the public
HEPData ratio should not be interpreted as a detector-level
validation of the comparison between mWW and effective QRA. Together, the two results
show that the expected yield need not in general factorize into a
single production rate normalization and a model-independent acceptance.
The public information is sufficient for the factorization test
performed here but does not provide a complete detector response for
arbitrary production models. We therefore do not construct a modified
NA62 exclusion. Equation~\eqref{eq:acceptance_shift} shows that the nonfactorization is
governed by the projection of the normalized production shape
variation onto the acceptance response. Such model dependence must
therefore be retained in the acceptance calculation or be treated as
an acceptance uncertainty rather than compressed into a single
normalization factor.

\section{Conclusions}
\label{sec:conclusion}

Normalizing the mWW and effective QRA spectra on a common physical
domain before applying the same acceptance response isolates a
substantial shape-induced acceptance difference.
 In the displayed BC1 overlap, all retained grid points satisfy
$\delta_{\rm NF}>0.3$ for the central calculation. The scan over $\Lambda_p=1.0$, $1.5$, and $2.0~\mathrm{GeV}$,
the variations of $R(Z)$, and the convergence check for the production
phase-space integration leave this conclusion unchanged. The observed nonfactorization is therefore not tied to a single
value of $\Lambda_p$ or a single aperture profile and occurs in parameter space overlapping the displayed NA62 BC1 sensitivity region.

The public NA62 full selection acceptances provide a complementary test at
the selection level. Over the analyzed HEPData window, the Mixing to
Bremsstrahlung dimuon ratio $F^{\rm sel}_{\rm NF,\mu\mu}$ ranges from
approximately 0.58 to 4.23, with a median absolute departure from unity of
approximately 0.79. 
This observable is distinct from the fiducial comparison between mWW
and effective QRA and does not constitute a detector-level validation
of that comparison. 

The publicly available information is insufficient to construct a modified
NA62 exclusion. The fiducial comparison should therefore not be interpreted
as a detector-level recast of the published limit. This limitation does not
affect the factorization test itself, which compares normalized production
shapes under a common acceptance response. For any specified response, a rate
factorization treatment is reliable only when the normalized production shape
variation has negligible projection onto that response at the required
accuracy. Otherwise, the production model dependence must be retained in the
acceptance calculation or treated as an acceptance uncertainty. The projection
criterion therefore provides a quantitative diagnostic for determining when
production model dependence can consistently be represented by a single rate
normalization.

\acknowledgments

This study was supported by the Fujian Province Young and Middle-aged Scientists Fund (grant JAT241139), the Fujian Province Natural Science Fund General Project (grant 2025J011710), and the Longyan University Doctoral Research Startup Projects 2025 (grant LB2025003). 

\section*{Data Availability}

The public NA62 dataset analyzed in this work was released by the
NA62 Collaboration and is available through HEPData~\cite{NA62_HEPData}. The numerical data generated in this study and supporting its findings
are available from the authors upon reasonable request.

\appendix

\setcounter{figure}{0}
\renewcommand{\thefigure}{B\arabic{figure}}

\section{Numerical Implementation}
\label{sec:numerical_implementation}

The mWW and effective QRA spectra are evaluated using the
corresponding expressions in Ref.~\cite{r22} together
with the timelike proton electromagnetic form factors from
the Unitary and Analytic model. Both spectra are normalized over
the same physical domain, with $0.01\leq z\leq0.99$ and
$p_T^2\leq1~\mathrm{GeV}^2$. Here $z$ denotes the longitudinal momentum fraction of the incident
proton carried by the dark photon, $p_T$ its transverse momentum with
respect to the beam axis, and $P_{\rm beam}=400~\mathrm{GeV}$ the
incident proton beam momentum.

The effective QRA hierarchy conditions are implemented numerically using an operational threshold of 0.20, requiring
\begin{equation}
\begin{aligned}
\frac{H}{4z(1-z)^2P_{\rm beam}^2}\leq 0.20,
\qquad\qquad&
p_T\leq 0.20P_{\rm beam},
\\[0.5em]
m_p\leq 0.20P_{\rm beam},
\qquad\qquad&
m_{A'}\leq 0.20E_{A'},
\end{aligned}
\label{eq:qra-validity-mask}
\end{equation}
where $m_p$ is the proton mass,
$H=p_T^2+z^2m_p^2+(1-z)m_{A'}^2$, and the high energy
approximation $E_{A'}\simeq zP_{\rm beam}$ is used in the numerical
implementation. These conditions define the common hierarchy mask
applied to both production models before normalization.

In the numerical implementation, Eq.~\eqref{eq:acceptance_kernel} is evaluated using the
longitudinal decay coordinate $Z$, for which
\begin{equation}
\frac{dP}{dZ}
=
\frac{1}{\lambda_Z}
\exp\!\left[-\frac{Z-Z_{\rm TAX}}{\lambda_Z}\right],
\qquad
\lambda_Z=\frac{p_L}{m_{A'}}c\tau,
\end{equation}
where $p_L$ and $p$ denote the longitudinal component and magnitude
of the dark photon momentum, respectively, and $Z_{\rm TAX}$ is the
longitudinal coordinate of the TAX reference position. This form
includes the path length Jacobian $ds/dZ=p/p_L$. The same prescription is used for both production
models. The hierarchy mask is applied within the production phase-space
integration for each mass point. 

The effective QRA weight contains the squared off-shell
factor $K_{\rm off}^2$, defined together with the auxiliary
quantity $\Delta_{\rm off}$ by
\begin{equation}
\begin{gathered}
K_{\rm off}
=
\frac{\Lambda_p^4}
{\Lambda_p^4+\Delta_{\rm off}^2},
\qquad
\Delta_{\rm off}=-\frac{H}{z}.
\end{gathered}
\label{eq:offshell-factor}
\end{equation}
The central calculation uses $\Lambda_p=1.5~\mathrm{GeV}$, with
$\Lambda_p=1.0$ and $2.0~\mathrm{GeV}$ taken as variations.

Both production models are passed through the same geometry and decay
acceptance kernel. The aperture $R(Z)$ is represented by a
piecewise constant profile reconstructed from the published NA62
detector geometry. The robustness test rescales the full profile by
$\pm10\%$. Convergence of the production phase-space integration is
checked against a lower-resolution $(z,p_T^2)$ grid, and points with
$|F_{\rm NF}^{\rm high}-F_{\rm NF}^{\rm low}|>0.03$
are excluded.

\section{Public NA62 Full Selection Acceptance Test}
\label{sec:OF_appendix}

\begin{figure}[t]
\includegraphics[width=0.86\columnwidth]{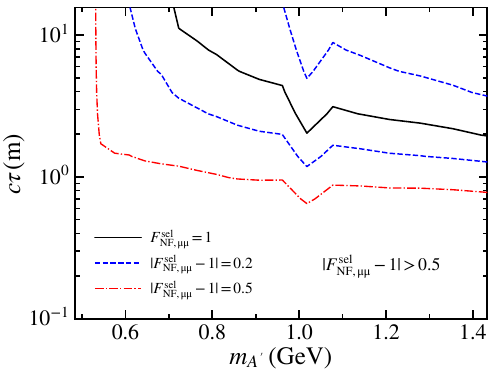}
\caption{
Public NA62 HEPData dimuon full selection acceptance ratio
$F^{\rm sel}_{\rm NF,\mu\mu}$.
Black solid, blue dashed, and red dash dotted contours denote
$F^{\rm sel}_{\rm NF,\mu\mu}=1$,
$|F^{\rm sel}_{\rm NF,\mu\mu}-1|=0.2$, and $0.5$,
respectively.
The labeled lower region has
$|F^{\rm sel}_{\rm NF,\mu\mu}-1|>0.5$.
Contours are evaluated directly on the published grid
without dense interpolation or smoothing.
}
\label{fig:fullselection}
\end{figure}

For the ratio defined in Eq.~\eqref{eq:fullselection-ratio}, the public
NA62 HEPData acceptance tables for the Mixing and Bremsstrahlung
production categories share the same mass and width grid, so the ratio is
formed point by point. The published acceptances are conditional on
the dark photon reaching the fiducial volume and decaying therein.
They therefore probe the reconstruction and analysis response after
this fiducial condition has been imposed, rather than the geometry and
decay probability isolated in the fiducial calculation.

The tabulated width is converted to $c\tau$ using $c\tau=\hbar c/\Gamma_{A'}$. Entries with vanishing or
undefined Bremsstrahlung acceptance are omitted before the ratio is
formed. No dense interpolation, smoothing, or resampling is applied
to the acceptance ratio. The contours in Fig.~\ref{fig:fullselection}
are evaluated directly on the published grid. 
Because the absolute
departure from unity is shown, the contour
$|F^{\rm sel}_{\rm NF,\mu\mu}-1|=0.2$ contains both the
$F^{\rm sel}_{\rm NF,\mu\mu}=0.8$ and
$F^{\rm sel}_{\rm NF,\mu\mu}=1.2$ branches.

Over the analyzed HEPData window,
$F^{\rm sel}_{\rm NF,\mu\mu}$ ranges from approximately
$0.58$ to $4.23$, with
$\operatorname{median}\left|F^{\rm sel}_{\rm NF,\mu\mu}-1\right|
\simeq 0.79$.
The departure from unity extends across neighboring grid points over
broad portions of the sampled mass and lifetime plane rather than
being confined to isolated points. The nonconstant ratio therefore
provides direct evidence from the published NA62 data that the full
selection acceptance depends on the production category at fixed
visible final state. 
This observable does not provide a detector-level validation of the mWW
to effective QRA comparison and is not used to modify the published NA62
exclusion.





\bibliographystyle{JHEP} 

\appendix

\end{document}